\overfullrule=0pt
\input harvmac
\def\a{{\alpha}}
\def\au{{\underline a}}
\def\ua{{\underline\alpha}}
\def\ad{{\dot \a}}
\def\adh{{\widehat{\dot \a}}}
\def\ah{{\widehat \a}}
\def\bd{{\dot \b}}
\def\bu{{\underline b}}
\def\cu{{\underline c}}
\def\bdh{{\widehat{\dot \b}}}
\def\l{{\lambda}}
\def\lh{{\widehat\lambda}}

\def\b{{\beta}}
\def\ub{{\underline\beta}}
\def\bh{{\widehat\beta}}
\def\dh{{\widehat\delta}}
\def\g{{\gamma}}
\def\ug{{\underline\gamma}}
\def\um{{\underline m}}
\def\gd{{\dot\gamma}}
\def\gdh{{\widehat{\dot \g}}}
\def\gh{{\widehat\gamma}}

\def\d{{\delta}}
\def\ud{{\underline\delta}}
\def\dd{{\dot\delta}}
\def\dh{{\widehat\d}}
\def\ddh{{\widehat{\dot \d}}}
\def\e{{\epsilon}}
\def\s{{\sigma}}

\def\half{{1\over 2}}
\def\p{{\partial}}

\def\t{{\theta}}
\def\th{{\widehat\theta}}

\def\bar{\overline}

\Title{\vbox{\hbox{IFT-P.041/2006 }}}
{\vbox{
\centerline{\bf Explaining Pure Spinor Superspace}}}
\bigskip\centerline{Nathan Berkovits\foot{e-mail: nberkovi@ift.unesp.br}}
\bigskip
\centerline{\it Instituto de F\'\i sica Te\'orica, State University of
S\~ao Paulo}
\centerline{\it Rua Pamplona 145, 01405-900, S\~ao Paulo, SP, Brasil}

\vskip .3in
In the pure spinor formalism for the superstring and supermembrane, 
supersymmetric invariants are constructed by
integrating over five $\theta$'s in d=10 and over nine $\theta$'s in d=11.
This pure spinor superspace is easily explained using the superform
(or ``ectoplasm'') method developed by Gates and collaborators, and
generalizes the standard chiral superspace in d=4. The ectoplasm method
is also useful for constructing d=10 and d=11 supersymmetric
invariants in curved
supergravity backgrounds.

\vskip .3in

\Date {November 2006}

\newsec{Introduction}

The conventional method for constructing supersymmetric invariants is
to integrate superfields over a superspace which contains
both $x^m$ and $\t^\a$
variables. The number of $\t$'s which must be integrated 
depends both on the spacetime dimension and on the constraints satisfied
by the superfields. For example, d=4 supersymmetric invariants can
be constructed either by using real superfields and integrating over
four $\t$'s, or by using chiral superfields and integrating over two
$\t$'s. Although it is somewhat non-trivial to generalize these
d=4 supersymmetric invariants in a curved supergravity background,
this can be done by inserting the appropriate constrained supervielbeins 
into the superspace integral.

This conventional method for constructing supersymmetric invariants
is less useful in higher spacetime dimensions which involve more $\t$'s.
For example, the construction of d=10 super-Poincar\'e invariant
expressions using unconstrained superfields would require integration
over 16 $\t$'s, which means that the supersymmetric invariants 
typically involve terms with eight spacetime derivatives. Although
one can try to define constrained d=10 superfields which allow integration
over fewer than 16 $\t$'s, finding an appropriate set of constraints
is not easy. Furthermore, if one finds a suitable set of constraints,
it is not obvious how to generalize them in a curved supergravity
background. 

Over the last six years, an appropriate set of constraints for 
d=10 and d=11 superfields has been discovered using the pure spinor
formalism for the superstring and supermembrane \ref\pureone
{N. Berkovits, {\it Super-Poincare covariant 
quantization of the superstring}, 
JHEP 0004 (2000) 018, hep-th/0001035.}\ref\puremem{N. Berkovits,
{\it Towards covariant quantization of the supermembrane},
JHEP 0209 (2002) 051, hep-th/0201151.}. Using 
the constraints coming from these pure spinor formalisms, 
d=10 and d=11 supersymmetric invariants in a flat background have been
constructed involving as few as two spacetime derivatives.
These supersymmetric invariants naturally arise as on-shell 
scattering amplitudes in the pure spinor approach.

For example, N=1 d=10 supersymmetric invariants can be constructed from
a superfield $f_{\a_1\a_2 \a_3}(x,\t)$ which satisfies the
constraint 
\eqn\onef{\l^\b\l^{\a_1} \l^{\a_2}\l^{\a_3} D_\b f_{\a_1 \a_2 \a_3}(x,\t)
=0}
where $\a=1$ to 16 is a d=10 spinor index,
$D_\a = {\p\over{\p\t^\a}}+\g^m_{\a\b}\t^\b$ 
is the superspace derivative, and
$\l^\a$ is a bosonic spinor satisfying the pure spinor condition that
$\l\g^m\l=0$ for $m=0$ to 9. 
The N=1 d=10 supersymmetric invariant is then obtained
by integrating over five of the 16 $\t$'s as 
\eqn\intonef{
T^{((\a_1 \a_2 \a_3))[\d_1 ... \d_5]} 
\int d^{10} x \int (d^5 \t)_{\d_1 ... \d_5}
~f_{\a_1\a_2\a_3} (x,\t)}
where 
\eqn\defT{
T^{\a_1 \a_2 \a_3~\d_1 ... \d_5} = \g_m^{\a_1 \d_1} \g_n^{\a_2 \d_2} 
\g_p^{\a_3 \d_3} (\g^{mnp})^{\d_4 \d_5}}
and 
$T^{((\a_1 \a_2 \a_3))[\d_1 ... \d_5]} $ is obtained from \defT\
by antisymmetrizing in the $\d$ indices, symmetrizing in the $\a$
indices, and subtracting $\g$-matrix trace terms in the $\a$ indices
so that 
$\g^m_{\a_1 \a_2} 
T^{((\a_1 \a_2 \a_3))[\d_1 ... \d_5]} =0 $. 
For example, the cubic d=10 super-Yang-Mills coupling is given by
\intonef\ where $f_{\a_1\a_2\a_3}= A_{\a_1} A_{\a_2} A_{\a_3}$
and $A_\a(x,\t)$ is the on-shell super-Yang-Mills spinor gauge superfield
\pureone.

One can also construct N=2 d=10 supersymmetric invariants from
a superfield
$f_{\a_1\a_2 \a_3 \bh_1\bh_2\bh_3}(x,\t, \th)$ which satisfies the
constraints
\eqn\twof{\l^\g\l^{\a_1} \l^{\a_2}\l^{\a_3}
\lh^{\bh_1} \lh^{\bh_2}\lh^{\bh_3}
 D_\g f_{\a_1 \a_2 \a_3
\bh_1\bh_2\bh_3}(x,\t,\th)=0, }
$$
\lh^\gh\lh^{\bh_1} \lh^{\bh_2}\lh^{\bh_3}
\l^{\a_1} \l^{\a_2}\l^{\a_3}
 D_\gh f_{\a_1 \a_2 \a_3
\bh_1\bh_2\bh_3}(x,\t,\th)=0, $$
where
$\a=1$ to 16 and $\bh=1$ to 16 are d=10 spinor indices which
are either of opposite chirality (for N=2A) or of the same
chirality (for N=2B),
$D_\a = {\p\over{\p\t^\a}}+\g^m_{\a\b}\t^\b$ and
$D_\ah = {\p\over{\p\th^\ah}}+\g^m_{\ah\bh}\th^\bh$ are the N=2
superspace derivatives, and
$\l^\a$ and $\lh^\bh$ are bosonic spinors satisfying the pure spinor
conditions that $\l\g^m\l=\lh\g^m \lh=0$. The N=2 d=10 supersymmetric
invariant is then obtained
by integrating over five $\t$'s and five $\th$'s as
\eqn\inttwof{
T^{((\a_1 \a_2 \a_3))[\d_1 ... \d_5]}
T^{((\bh_1 \bh_2 \bh_3))[\gh_1 ... \gh_5]}
\int d^{10} x \int (d^5 \t)_{\d_1 ... \d_5}
\int (d^5 \th)_{\gh_1 ... \gh_5} ~
f_{\a_1\a_2\a_3 \bh_1 \bh_2\bh_3}. }

Finally, d=11 supersymmetric invariants can be constructed from
a superfield 
$f_{\ua_1... \ua_7}(x,\t)$ which satisfies the
constraint
\eqn\threef{\l^\ug\l^{\ua_1} ...\l^{\ua_7} D_\ug f_{\ua_1 ... \ua_7}
(x,\t)=0}
where $\ua=1$ to 32 is a d=11 spinor index, 
$D_\ua = {\p\over{\p\t^\ua}}+\g^\um_{\ua\ub}\t^\ub$ 
is the d=11 superspace derivative, 
and $\l^\ua$ is a bosonic spinor satisfying the condition that
$\l\g^{\underline m}\l=0$ for $\um=0$ to 10.
The d=11 supersymmetric invariant is obtained by integrating over
9 of the 32 $\t$'s as 
\eqn\intthreef{
T^{((\ua_1 ...\ua_7))[\ud_1 ... \ud_9]}
\int d^{11} x \int (d^9 \t)_{\ud_1 ... \ud_9} ~
f_{\ua_1 ... \ua_7}(x,\t). }
As in d=10, 
$T^{((\ua_1 ... \ua_7))[\ud_1 ... \ud_9]}$ is
a Lorentz-invariant tensor which is antisymmetric in the $\ud$
indices and symmetric $\g$-matrix traceless in the $\ua$ indices.
The explicit expression for
$T^{((\ua_1 ... \ua_7))[\ud_1 ... \ud_9]}$ in terms of $\g$-matrices
is a bit
more complicated than in d=10, however, it
can be defined indirectly through
the formula
\eqn\defTT{(\l\g^{\um_1}\t) ... (\l\g^{\um_9}\t) =
T^{((\ua_1 ... \ua_7))[\ud_1 ... \ud_9]}
\l_{\ua_1} ... \l_{\ua_7}
\t_{\ud_1} ... \t_{\ud_9}
~(\l \g^{\um_1 ... \um_9} \l)}
for any $[\um_1 ... \um_9]$. 

The d=10 and d=11 supersymmetric invariants of
\intonef, \inttwof\ and \intthreef\
were originally constructed
by looking for elements of top ghost number in the pure
spinor BRST cohomology \ref\purehowe{P.S. Howe, 
{\it Pure spinors lines in superspace and ten-dimensional supersymmetric
theories,} Phys. Lett.  B258 (1991) 141\semi
P.S. Howe,
{\it Pure spinors, function superspaces and supergravity theories in
  ten-dimensions and eleven-dimensions,} Phys. Lett.  B273 (1991) 
90.}\pureone\puremem.
Using the N=1 d=10 nilpotent BRST operator $Q_{N=1} = \l^\a D_\a$ \purehowe,
the top element in the BRST cohomology is\pureone
\eqn\topone{(\l\g^m\t)(\l\g^n\t)(\l\g^p\t)(\t\g_{mnp}\t).}
Since \topone\ cannot be written as the supersymmetric variation of
a BRST-closed operator, and since \intonef\ selects out the component
of $\l^\a\l^\b\l^\g f_{\a\b\g}$ proportional to \topone, \intonef\
is supersymmetric if 
$\l^\a\l^\b\l^\g f_{\a\b\g}$ is BRST closed, i.e. if $f_{\a\b\g}$
satisfies \onef. Similarly, the top element in the cohomology of the
N=2 d=10 BRST operator $Q_{N=2} = \l^\a D_\a + \lh^\ah D_\ah$ is
\eqn\toptwo{(\l\g^{m_1}\t)(\l\g^{m_2}\t)(\l\g^{m_3}\t)(\t\g_{m_1 m_2 m_3}\t)
~(\lh\g^{n_1}\th)(\lh\g^{n_2}\th)(\lh\g^{n_3}\th)(\th\g_{n_1 n_2 n_3}\th), }
and the top element in the cohomology of the d=11 BRST operator
$Q_{d=11} = \l^\ua D_\ua$ is\puremem
\eqn\topthree{ \l_{\ua_1} ... \l_{\ua_7}
T^{((\ua_1 ... \ua_7))[\ud_1 ... \ud_9]}\t_{\ud_1} ... \t_{\ud_9}.}
So \inttwof\ and \intthreef\ are supersymmetric if
\eqn\brstc{Q_{N=2}(\l^\a \l^\b \l^\g \lh^\ah\lh^\bh\lh^\gh
f_{\a\b\g~\ah\bh\gh}) =0 \quad {\rm and} \quad
Q_{d=11}(\l^{\ua_1}... \l^{\ua_7}
f_{\ua_1 ... \ua_7}) =0, }
i.e. if \twof\ and \threef\ are satisfied.

Although this pure spinor construction is hard to understand using  
the conventional method for constructing supersymmetric invariants,
it will be easy to explain this construction using the superform
(or ``ectoplasm'') method developed by Gates and collaborators \ref\gates
{S.J. Gates, Jr., {\it Ectoplasm has no topology}, Nucl. Phys. B541 (1999)
615, hep-th/9809056\semi
S.J. Gates, Jr., {\it Ectoplasm has no topology: the prelude}, 
hep-th/9709104.}\ref\etall{S.J. Gates, Jr., M. Grisaru, M. Knutt-Wehlau
and W. Siegel, {\it Component actions from
curved superspace: normal coordinates and ectoplasm}, Phys. Lett.
B421 (1998) 203, hep-th/9711151.}
for
constructing supersymmetric invariants. The superform (or ``ectoplasm'')
method will also be useful for generalizing these d=10 and d=11
invariants in a curved supergravity background. 

When constructed using the superform method, the invariants of
\intonef, \inttwof\ and \intthreef\ will turn out to be natural
d=10 and d=11 generalizations of chiral superspace
integrals in four dimensions. This is not surprising since, as was shown
in \ref\twopure{N. Berkovits, {\it Pure spinor formalism as an N=2
topological string}, JHEP 01510 (2005) 089, hep-th/0509120.}, there exists
a four-dimensional version of the pure spinor formalism whose
scattering amplitudes compute chiral F-terms in the 
d=4 effective action.

In section 2 of this paper, the superform method for constructing 
N=1 and N=2 supersymmetric invariants in four dimensions will be reviewed.
And in section 3, the superform method will be used to construct
the N=1 d=10, N=2A d=10, and d=11 supersymmetric invariants of 
\intonef, \inttwof\ and \intthreef. Surprisingly, the N=2B d=10
supersymmetric invariant of \inttwof\ does not 
have an obvious construction using the superform method.\foot
{Some of the results in this paper on N=1 d=10 and N=1
d=11 invariants were discussed in a talk given by Paul Howe in January 2005
\ref\howetalk{P.S. Howe, ``Pure spinors and spinorial cohomology'',
Talk given at the IHES/CEA Workshop on the Pure Spinor Formalism
in String Theory, Bures-sur-Yvette, France, January 2005.}.
The contents of this talk, together with later developments, can
be found in \ref\newpaper{N. Berkovits and P.S. Howe,
{\it The cohomology of superspace, pure spinors and invariant
integrals}, hep-th/0803.3024.}.}

\newsec{Review of Superform (or ``Ectoplasm'') Method}

The superform (or ``ectoplasm'')
method was developed in papers by Gates \gates\ and by
Gates, Grisaru, Knutt-Wehlau and Siegel \etall, and has connections with
work on ``rheonomy'' \ref\rheon{R. D'Auria, P. Fre, P.K. Townsend
and P. van Nieuwenhuizen, {\it Invariance of actions, rheonomy and the
new minimal N=1 supergravity in the group manifold approach},
Annals Phys. 155 (1984) 423.}\ref\grassifre{
P. Fre and P.A. Grassi,
{\it Pure spinors, free differential algebras, and the supermembrane},
  Nucl. Phys. B763 (2007) 1, 
hep-th/0606171.}
and brane embeddings \ref\howeemb
{P.S. Howe, O. Raetzel and E. Sezgin, 
{\it On brane actions and superembeddings},
JHEP 9808 (1998) 011, hep-th/9804051.}\ref\sortwo{
I.A. Bandos, D.P. Sorokin and D. Volkov,
{\it On the generalized action principle for superstrings and supermembranes},
Phys. Lett. B352 (1995) 269,
hep-th/9502141.}.
The superform method has previously been used to reproduce supergravity
actions \gates\etall, to construct 
new supersymmetric
invariants in three \ref\ectthree{J.M. Drummond, P. Howe and
U. Lindstrom, {\it
Kappa symmetric non-abelian Born-Infeld actions in
three dimensions}, Class. Quant. Grav. 19 (2002) 6477, hep-th/0206148\semi
P. Howe and
U. Lindstrom, {\it 
Kappa symmetric higher derivative terms in brane actions},
Class. Quant. Grav. 19 (2002) 2813, hep-th/0111036\semi
M. Becker, D. Constantin, S.J. Gates, Jr., W. Linch III, W. Merrell
and J. Phillips, {\it M-theory on Spin(7) manifolds, fluxes and 3D N=1
supergravity},  Nucl. Phys. B683 (2004) 67, hep-th/0312040.}, 
four \ref\ectfour{T. Biswas and W. Siegel, {\it N=2 harmonic superforms,
multiplets and actions}, JHEP 0111 (2001) 004, hep-th/0105084.} and six
\ref\ectsix{J.M. Drummond and P.S. Howe, {\it Codimension zero 
superembeddings}, Class. Quant. Grav. 18 (2001) 4477, hep-th/0103191.}
dimensions, and to construct supersymmetric Chern-Simons
terms in any dimension \ref\ectcs{P.S. Howe and D. Tsimpis, {\it On
higher order corrections in M 
theory}, JHEP 0309 (2003) 038, hep-th/0305129.}.
The relation between the
superform method and
the pure spinor constructions has some
similarities with the 
superaction formalism of \ref\superaction{P. Howe, K. Stelle and
P. Townsend, {\it Superactions}, Nucl. Phys. B191 (1981) 445.} and with the
relation found by
Cederwall, Nilsson and Tsimpis \ref\ceder{M. Cederwall, B.E.W. Nilsson
and D. Tsimpis, {\it The structure of maximally supersymmetric Yang-Mills
theory: constraining higher order corrections}, JHEP 0106 (2001) 034,
hep-th/0102009\semi
M. Cederwall, B.E.W. Nilsson
and D. Tsimpis, {\it Spinorial cohomology and
maximally supersymmetric theories}, JHEP 0202 (2002) 009, hep-th/0110069.}
between maximally supersymmetric deformations and
spinorial cohomology. 

The basic idea of the superform method is to look for
a closed superform $J_{M_1 ... M_d}(x, \t)$ where $d$ is the
dimension of spacetime and $M=(m,\mu)$ is either a spacetime vector
index $m$ or a spacetime spinor index $\mu$. Note that
superforms are graded-antisymmetric, i.e. they are antisymmetric
in the vector indices and symmetric in
the spinor indices. In terms of 
$J_{M_1 ... M_d}(x,\t)$,
the supersymmetric invariant is given simply by
\eqn\superi{I = {1\over{d!}}
\e^{m_1 ... m_d}\int d^d x  ~J_{m_1 ... m_d}(x,\t=0).}
When $J_{M_1 ... M_d}$ is closed (i.e. $ \p_{[N} J_{M_1 ... M_d)}=0$
where $[~~)$ denotes commutator for vector indices and anticommutator
for spinor indices),
$I$ is supersymmetric since 
\eqn\checki{\int d^d x  ~ {\p\over{\p\t^\mu}}J_{m_1 ... m_d}(x,\t)
= 
{{(-1)^{d+1}}\over{(d-1)!}}
\int d^d x  ~ \p_{[m_1} J_{m_2 ... m_d] \mu}(x,\t) =0}
if one ignores surface terms. 

Furthermore, this method is easily generalized to a curved
supergravity background by defining 
\eqn\supertwoi{I = {1\over{d!}}\e^{m_1 ... m_d}\int d^d x  ~
e_{m_d}^{A_d}(x) ... e_{m_1}^{A_1}(x)~J_{A_1 ... A_d}(x,\t=0)}
where $A=(a,\a)$ are tangent-superspace indices, $e_m^a(x)$ and
$e_m^\a(x)$ are the vielbein and gravitino, 
$J_{A_1 ... A_d}(x,\t)$
is a covariantly closed $d$-superform satisfying
\eqn\covclosed{ D_{[B} J_{A_1 ... A_d)} = {d\over 2}
~T_{[B A_1|}{}^C ~J_{C| A_2 ... A_d)},}
and $T_{AB}{}^C$ is the supertorsion.
The formula of \covclosed\ can be derived from the relation
\eqn\covj{E^{A_d} ... E^{A_1} J_{A_1 ... A_d} =
dZ^{M_d} ... dZ^{M_1} J_{M_1 ... M_d}}
where $E^A = dZ^M E_M^A$ is the vielbein superform, $E_M^A$
is the supervielbein, $dZ^M = (dx^m, d\t^\mu)$, and
$\p_{[N} J_{M_1 ... M_d)} =0.$
Note that $I$ of \supertwoi\ is invariant under the gauge transformation
\eqn\gauget{\d J_{A_1 ... A_d} = {1\over {(d-1)!}}
D_{[A_1} \Lambda_{A_2 ... A_d)} - {1\over{2(d-2)!}}
T_{[A_1 A_2|}{}^C \Lambda_{C| A_3 ... A_d)}}
since under \gauget, 
$\d J_{m_1 ... m_d} = {1\over{(d-1)!}} \p_{[m_1}\Lambda_{m_2 ... m_d)}$.

Solving \covclosed\ for $J_{A_1 ... A_d}$
only requires knowledge of the supertorsion and supercurvature, so
the explicit superfield for the supervielbein is unnecessary for
constructing the supersymmetric invariant of \supertwoi.
Since the supervielbein is usually a complicated
superfield, this is a big advantage over the conventional approach
to constructing supersymmetric invariants in a curved background.

In looking for solutions to \covclosed\ in a flat background where
the only non-zero torsion is
$T_{\a\b}{}^c= \g_{\a\b}^c$, it will turn out that $J_{a_1 ... a_d}(x,\t)$
with all vector indices can be related to 
$J_{a_1 ... a_{d-N} \b_1 ... \b_N}(x,\t)$
with $d-N$ vector indices by acting with $N$ spinor derivatives, i.e.
\eqn\relv{J_{a_1 ... a_d}(x,\t) = D_{\g_1} ... D_{\g_N}
J_{a_1 ... a_{d-N} \b_1 ... \b_N}(x,\t)}
where the index contractions on the right-hand side of \relv\
need to be worked out. Furthermore, one finds that
when $N$ is larger than some fixed
value $L$, $J_{a_1 ... a_{d-N} \b_1 ... \b_N}(x,\t)=0$.
So in a flat background, the supersymmetric invariant can be written as
\eqn\superthreei{I = {1\over {d!}}\e^{a_1 ... a_d}\int d^d x  ~
J_{a_1 ... a_d}(x,\t=0) =
\int d^d x  ~
D_{\g_1} ... D_{\g_L}
J_{a_1 ... a_{d-L} \b_1 ... \b_L}(x,\t=0)}
$$= 
\int d^d x  \int (d^L \t)_{\g_1 ... \g_L}
~J_{a_1 ... a_{d-L} \b_1 ... \b_L}(x,\t)$$
for some contraction of the spinor and vector indices.
Determining the conditions for $J_{a_1 ... a_{d-L} \b_1 ... \b_L}(x,\t)$
to satisfy \covclosed\ 
is equivalent in the conventional approach to finding the appropriate
set of constraints for the superfields which allow integration over
$L$ $\t$'s. 

\subsec{N=1 d=4 invariants}

To reproduce the standard N=1
d=4 chiral superspace integral using the superform method, 
one imposes that the maximum number of spinor indices on 
$J_{A_1 ... A_4}(x,\t)$ is two and that 
\ref\gatesold{S.J. Gates, Jr., 
{\it Super p-form gauge superfields}, Nucl. Phys.  B184 (1981) 381.}
\eqn\onefour{J_{ab\g\d}(x,\t) = (\g_{ab})_{\g\d} \bar V(x,\t), \quad
J_{ab\gd\dd}(x,\t) = (\g_{ab})_{\gd\dd} V(x,\t), }
where $a=0$ to 3 are vector indices,
$\a=1$ to 2 and $\ad=1$ to 2 are Weyl and anti-Weyl spinor indices,
$V$ and $\bar V$ are chiral and antichiral superfields satisfying
$D_\gd V=0$ and $D_\g \bar V=0$, and $(\g_{ab})_{\g\d}$ and
$(\g_{ab})_{\gd\dd}$ are the self-dual and anti-self-dual two-form
$\g$-matrices.

In a flat background, the only non-zero torsion is $T_{\a\bd}{}^c=
\s_{\a\bd}^c$ and the chirality conditions on $V$ and $\bar V$ come
from the constraints that 
$D_{(\a} J_{\b\g ) ab}=0$ and
$D_{(\ad} J_{\bd\gd ) ab}=0$.
Furthermore, the gauge parameter $\Lambda_{abc}$ of \gauget\ can be used
to gauge $J_{ab \a \bd}=0$. 
The constraints $D_{\ad} J_{ab\b\g}=  T_{\ad(\b}{}^c J_{\g)abc}$ and
$D_{\a} J_{ab\bd\gd}= T_{\a(\bd}{}^c J_{\gd)abc}$ imply that
$J_{abc\g} = \e_{abcd} \s^d_{\g\bd} D^\bd \bar V$ and
$J_{abc\gd} = \e_{abcd} \s^d_{\b\gd} D^\b V$. And the constraint
$D_{(\ad} J_{\a) abc} =  T_{\a\ad}{}^d J_{dabc}$ implies that
$J_{abcd} = \e_{abcd}(D_\a D^\a V + D_\ad D^\ad \bar V)$.
So the supersymmetric invariant is 
\eqn\invone{I = {1\over {4!}}\e^{abcd}\int d^4 x J_{abcd} =
\int d^4 x (D_\a D^\a V + D_\ad D^\ad \bar V),}
which reproduces the standard d=4 chiral superspace integral
$I = \int d^4 x (\int d^2 \t ~V + \int d^2 \bar\t ~ \bar V)$.

\subsec{N=2 d=4 invariants}

For the N=2 d=4 case, one finds that the maximum number of
spinor indices on $J_{A_1 ... A_4}(x,\t,\th)$ is four, and that
\ectfour
\eqn\twofour{J_{\a\b\gh\dh}(x,\t,\th) = (\g_{ab})_{\a\b}(\g^{ab})_{\gh\dh}
\bar W(x,\t,\th), \quad
J_{\ad\bd\gdh\ddh}(x,\t,\th) = (\g_{ab})_{\ad\bd}(\g^{ab})_{\gdh\ddh}
W(x,\t,\th), }
where $\a,\bd,\gh,\ddh=1$ to 2,
$W$ and $\bar W$ are chiral and antichiral
superfields satisfying the constraints
$D_\gd W=D_\ddh W=0$
and
$D_\g \bar W=D_\dh \bar W=0$, and all other components of
$J_{A_1 ... A_4}$ with four spinor indices are zero.

In a flat background, the only non-zero torsions are
$T_{\a\bd}{}^c=
\s_{\a\bd}^c$ and
$T_{\ah\bdh}{}^c=
\s_{\ah\bdh}^c$, and the chirality conditions on $W$ and $\bar W$ come
from the constraints that
\eqn\constwo{D_{(\a} J_{\b\g ) \bh\gh}=
D_{(\ah} J_{\bh\gh ) \b\g}=
D_{(\ad} J_{\bd\gd ) \bdh\gdh}=
D_{(\adh} J_{\bdh\gdh ) \bd\gd}= 0.}
As in the N=1 d=4 case, the vector components of $J_{A_1 ... A_4}$
can be determined from the spinor components of \twofour\ using the
constraints of \covclosed. One finds that
$J_{abcd}= \e_{abcd} (D_\a D^\a D_\bh D^\bh W +
D_\ad D^\ad D_\bdh D^\bdh \bar W )$, so the supersymmetric invariant
\eqn\twofive{I ={1\over{4!}} \e^{abcd}\int d^4 x J_{abcd}
= \int d^4 x
(D_\a D^\a D_\bh D^\bh W +
D_\ad D^\ad D_\bdh D^\bdh \bar W )}
coincides with the standard N=2
chiral superspace integral
$I = \int d^4 x (\int d^2 \t d^2 \th~ W + \int d^2 \bar\t \int d^2 \widehat
{\bar\t} ~\bar W)$.

\newsec{Superform Method in Higher Dimensions}

\subsec{N=1 d=10 invariants} 

In any even spacetime dimension $d=2R$, 
there is a natural generalization of the
N=1 d=4 formula of \onefour\ for the superforms. 
The generalization is that
the maximum number of spinor indices of $J_{A_1 ... A_d}(x,\t)$ is 
$R={d\over 2}$ and that 
\eqn\highone{J_{a_1 ... a_R \b_1 ... \b_R}(x,\t) = 
(\g_{a_1 ... a_R})_{(\b_1 \b_2}~~ f_{\b_3 ... \b_R)}(x,\t)}
where $f_{\a_1 ... \a_{R-2}}(x,\t)$ is a superfield satisfying the constraint
\eqn\onehigh{\l^\b\l^{\a_1} ... \l^{\a_{R-2}} D_\b f_{\a_1 ... \a_{R-2}}=0,}
$(\g_{a_1 ... a_R})_{\b\g} =
(\g_{a_1 ... a_R})_{\g\b}$ 
is the self-dual ${d\over 2}$-form $\g$-matrix,
and
$\l^\a$ is a bosonic spinor satisfying the condition that
$\l\g^c\l=0$ for $c=0$ to ${d-1}$. 

To show that \highone\ satisfies \covclosed\
in a flat background where the only non-vanishing torsion is 
$T_{\a\b}{}^c = \g_{\a\b}^c$, note that \onehigh\ implies that
$ \l^\g\l^{\b_1} ... \l^{\b_R} D_\g J_{\b_1 ... \b_R a_1 ... a_R}=0$.
Since $\l\g^c\l=0$, this implies that
\eqn\checkonehigh{ D_{(\g} J_{\b_1 ... \b_R) a_1 ... a_R}=
\g^c_{(\g\b_1} K_{\b_2 ... \b_R)  a_1 ... a_R c}}
for some 
$K_{\b_2 ... \b_R a_1 ... a_R c}$. If one chooses
$J_{\b_1 ... \b_{R-1} a_1 ... a_{R+1}}$ such that it is proportional to
$K_{\b_1 ... \b_{R-1} a_1 ... a_{R+1} }$,
the first non-trivial constraint of \covclosed\ is satisfied.
Furthermore, the gauge invariance of \gauget\ implies that
$J_{a_1 ... a_R \b_1 ... \b_R}$ is defined up to the
gauge transformation
\eqn\gaugeonehigh{\d 
J_{a_1 ... a_R \b_1 ... \b_R} = {1\over{(R-1)!}}
D_{(\b_1} \Lambda_{\b_2 ... \b_R) a_1 ... a_R} -{1\over{2(R-2)!}}
  \g^c_{(\b_1 \b_2} \Lambda_{\b_3 ... \b_R)
a_1 ... a_R c}.}

As in the d=4 case, components of $J_{A_1 ... A_d}$ with more than
${d\over 2}$
vector components can be constructed from spinor
derivatives of $J_{a_1 ... a_R \b_1 ... \b_R}$
of \highone\ by using the constraints of \covclosed. In a flat background,
the supersymmetric invariant will therefore have the form
\eqn\superteni{I = {1\over {d!}}\e^{a_1 ... a_d}\int d^d x  ~
J_{a_1 ... a_d}(x,\t=0) =
\int d^d x  \int (d^R \t)_{\d_1 ... \d_R} 
~f_{\b_1 ... \b_{R-2}} (x,\t)}
where the index contractions need to be worked out.

When $d=10$, 
$J_{a_1 ... a_5 \b_1 ... \b_5}(x,\t) = 
(\g_{a_1 ... a_5})_{(\b_1 \b_2}~ f_{\b_3 \b_4\b_5)}(x,\t)$
where $f_{\a\b\g}$ satisfies the same constraints as in \onef.
To show that \superteni\ reproduces the supersymmetric invariant of
\intonef, note that the gauge invariance of \gaugeonehigh\ implies
that \superteni\ is invariant
under
\eqn\gaugeonefhigh{\d f_{\a\b\g } =  
\half D_{(\a}\Sigma_{\b\g)}
+ \half \g^c_{(\a \b} \Omega_{\g)c}}
where
\eqn\lambdao{
\Lambda_{\b_1\b_2\b_3\b_4}^{a_1 ... a_5} = {1\over {4}} 
(\g^{a_1 ... a_5})_{(\b_1\b_2} \Sigma_{\b_3\b_4)}, \quad
\Lambda_{\b_1\b_2\b_3}^{a_1 ... a_5 c} = {1\over {240}}
(\g^{[a_1 ... a_5})_{(\b_1\b_2}
\Omega^{c]}_{\b_3)}.}
In relating \gaugeonehigh\ and \gaugeonefhigh, one needs to use the
d=10 identity $(\g_c)_{(\b_1\b_2} (\g^{c a_1 a_2 a_3 a_4})_{\b_3\b_4)}=0$.

The gauge invariance of \gaugeonefhigh\ implies that \superteni\
only depends on the $\g$-matrix traceless part
of $f_{\a\b\g}$ and is invariant under
\eqn\gafone{\d (\l^\a\l^\b\l^\g f_{\a\b\g}) =  \l^\a D_\a
( \l^\b\l^\g \Sigma_{\b\g}) =  Q_{N=1} 
( \l^\b\l^\g \Sigma_{\b\g}).}
So \superteni\ is independent of BRST-trivial deformations of
$\l^\a\l^\b\l^\g f_{\a\b\g}$. Since the unique state with three
$\l$'s in the BRST cohomology is \topone, \superteni\ selects out
the component of 
$\l^\a\l^\b\l^\g f_{\a\b\g}$ proportional to \topone, and is therefore
proportional to 
\eqn\inthighonef{
T^{((\b_1 \b_2 \b_3))[\d_1 ... \d_5]} 
\int d^{10} x \int (d^5 \t)_{\d_1 ... \d_5}
~f_{\b_1\b_2\b_3} (x,\t)}
of \intonef.

To generalize this N=1 d=10 supersymmetric invariant in a curved
supergravity background, one first defines 
$J_{a_1 ... a_5 \b_1 ... \b_5}(x,\t) = 
(\g_{a_1 ... a_5})_{(\b_1 \b_2} 
f_{\b_3 ... \b_5)}(x,\t)$ as in \highone, but
where $D_\b$ of \onehigh\ is now the spinor derivative in a curved
background. One then needs to compute the other components of
$J_{A_1 ... A_{10}}$ in terms of $f_{\a\b\g}(x,\t)$ using the constraints
of \covclosed. Finally, one plugs the $\t=0$ components of 
$J_{A_1 ... A_{10}}$ into the supersymmetric invariant  
\eqn\superteni{I = {1\over{10!}}\e^{m_1 ... m_{10}}\int d^{10} x  ~
e_{m_{10}}^{A_{10}}(x) ... e_{m_{1}}^{A_{1}}(x)~J_{A_1 ... A_{10}}(x,\t=0)}
where $e_m^a(x)$ and $e_m^\a(x)$ are the ten-dimensional vielbein and
gravitino.

\subsec{N=2A d=10 invariants}

In any even spacetime dimension $d=2R$,
a natural generalization of the
N=2 d=4 formula of \twofour\
is that
the maximum number of spinor indices of $J_{A_1 ... A_d}(x,\t, \th)$ is
$d =2 R$ and that
\eqn\hightwo{J_{\a_1 ... \a_R \bh_1 ... \bh_R}(x,\t,\th) =
(\g_{c_1 ... c_R})_{(\a_1 \a_2}
~~ f_{\a_3 ... \a_R)(\bh_1 ... \bh_{R-2} }(x,\t,\th) ~
~(\g^{c_1 ... c_R})_{\bh_{R-1} \bh_{R})} }
where $f_{\a_1 ... \a_{R-2}\bh_1 ... \bh_{R-2}}(x,\t,\th)$
is a superfield satisfying the constraints
\eqn\twohigh{\l^\g\l^{\a_1} ... \l^{\a_{R-2}}\lh^{\bh_1} ... \lh^{\bh_{R-2}}
D_\g f_{\a_1 ... \a_{R-2}\bh_1 ... \bh_{R-2}}=0,}
$$\lh^\gh\lh^{\bh_1} ... \lh^{\bh_{R-2}}
\l^{\a_1} ... \l^{\a_{R-2}}
D_\gh f_{\a_1 ... \a_{R-2}\bh_1 ... \bh_{R-2}}=0,$$
and
$\l^\a$ and $\lh^\bh$ are bosonic spinors satisfying the conditions that
$\l\g^c\l= \lh\g^c\lh= 0$ for $c=0$ to ${d-1}$.

To show that \hightwo\ satisfies \covclosed\
in a flat background where the only non-vanishing torsions are
$T_{\a\b}{}^c = \g_{\a\b}^c$ and
$T_{\ah\bh}{}^c = \g_{\ah\bh}^c$,
note that \twohigh\ implies that
\eqn\impltwo{ \l^\g\l^{\a_1} ... \l^{\a_R} \lh^{\bh_1} ... \lh^{\bh_R}
D_\g J_{\a_1 ... \a_R \bh_1 ... \bh_R}=0, }
$$
\lh^\gh
\lh^{\bh_1} ... \lh^{\bh_R}
\l^{\a_1} ... \l^{\a_R}
D_\gh J_{\a_1 ... \a_R \bh_1 ... \bh_R}=0.$$ 
Since $\l\g^c\l=\lh\g^c\lh=0$, this implies that
\eqn\checktwohigh{ D_{(\g} J_{\a_1 ... \a_R) \bh_1 ... \bh_R}=
\g^c_{(\g\a_1} K_{\a_2 ... \a_R) c \bh_1 ... \bh_R} +
\g^c_{(\bh_1\bh_2} K_{\bh_3 ... \bh_R) c \g \a_1 ... \a_R}, }
$$D_{(\gh} J_{\bh_1 ... \bh_R) \a_1 ... \a_R}=
\g^c_{(\gh\bh_1} K_{\bh_2 ... \bh_R) c \a_1 ... \a_R} +
\g^c_{(\a_1\a_2} K_{\a_3 ... \a_R) c \gh \bh_1 ... \bh_R}, $$
for some choice of $K$'s with one vector index and $d-1$ spinor
indices. If one sets $J$'s with one vector index and $d-1$ spinor
indices to be proportional to these $K$'s, the first non-trivial
condition coming from \covclosed\ is satisfied.
Furthermore, the gauge invariance of \gauget\ implies that
$f_{\a_1 ... \a_{R-2}
\bh_1 ... \bh_{R-2} }$ is defined up to the gauge
transformation
\eqn\gaugetwofhigh{\d
f_{\a_1 ... \a_{R-2}\bh_1 ... \bh_{R-2} }
=  D_{(\a_1} \Sigma_{\a_2 ... \a_{R-2})\bh_1 ... \bh_{R-2}} +
 D_{(\bh_1} \widehat\Sigma_{\bh_2 ... \bh_{R-2})\a_1 ... \a_{R-2}} }
$$ +
\g^c_{(\a_1 \a_2} \Omega_{\a_3 ... \a_{R-2})\bh_1 ... \bh_{R-2} c}
+  \g^c_{(\bh_1 \bh_2} 
\widehat\Omega_{\bh_3 ... \bh_{R-2})\a_1 ... \a_{R-2} c} .$$

As in the N=2 d=4 case, components of $J_{A_1 ... A_d}$ with
vector components can be constructed from spinor
derivatives of $J_{\a_1 ... \a_R \bh_1 ... \bh_R}$
of \hightwo\ by using the constraints of \covclosed. In a flat background,
the supersymmetric invariant will therefore have the form
\eqn\supertenf{I = \e^{a_1 ... a_d}\int d^d x  ~
J_{a_1 ... a_d}(x,\t=\th=0)}
$$ =
\int d^d x  \int (d^R \t)_{\g_1 ... \g_R}
(d^R \th)_{\dh_1 ... \dh_R}
~f_{\a_1 ... \a_{R-2}\bh_1 ... \bh_{R-2}}$$
where the index contractions need to be worked out.

When $d=10$,
\eqn\dtwo{J_{\a_1 ... \a_5 \bh_1 ... \bh_5}(x,\t,\th) =
(\g_{c_1 ... c_5})_{(\a_1 \a_2}~~ f_{\a_3 \a_4\a_5)(\bh_1 \bh_2 \bh_3}
(x,\t,\th)
~~(\g^{c_1 ... c_5})_{\bh_4 \bh_5)}}
where $f_{\a_1\a_2\a_3\bh_1\bh_2 \bh_3}$
satisfies the same constraints as in \twof. However, since
\eqn\prodgg{(\g_{c_1 ... c_5})_{\a_1 \a_2}
~(\g^{c_1 ... c_5})_{\bh_1 \bh_2}}
 vanishes when $\a$ and $\bh$ are
d=10 spinors of the same chirality, \dtwo\ can only be used for the
N=2A case. So there is no obvious way to construct the
N=2B supersymmetric invariant of \inttwof\ using
the superform method.

To show that \supertenf\ reproduces the supersymmetric
invariant of \inttwof\ for the N=2A case,
note that \gaugetwofhigh\ implies that \supertenf\ is independent
of the $\g$-matrix traceless components of 
$f_{\a_1\a_2\a_3 \bh_1 \bh_2 \bh_3}$ and 
is invariant under BRST-trivial deformations of the form
\eqn\gaftwo{\d (\l^{\a_1}\l^{\a_2} \l^{\a_3}
\lh^{\bh_1}\lh^{\bh_2} \lh^{\bh_3}
f_{\a_1\a_2\a_3 \bh_1 \bh_2 \bh_3}) =}
$$ Q_{N=2} (
\l^{\a_2} \l^{\a_3}
\lh^{\bh_1}\lh^{\bh_2} \lh^{\bh_3}
\Sigma_{\a_2\a_3 \bh_1 \bh_2 \bh_3} +
\l^{\a_1}\l^{\a_2} \l^{\a_3}
\lh^{\bh_2} \lh^{\bh_3}
\widehat\Sigma_{\a_1\a_2\a_3  \bh_2 \bh_3} ).$$
Since the unique state with three
$\l$'s and three $\lh$'s
in the BRST cohomology is \toptwo, \supertenf\ selects out
the component of 
$\l^{\a_1}\l^{\a_2} \l^{\a_3}
\lh^{\bh_1}\lh^{\bh_2} \lh^{\bh_3}
f_{\a_1\a_2\a_3 \bh_1 \bh_2 \bh_3}$
proportional to \toptwo, and therefore
reproduces the supersymmetric invariant of \inttwof.

\subsec{d=11 invariants} 

Finally, it will be shown how to
construct the d=11 supersymmetric invariant of \intthreef\
using the superform method. Although there is no obvious
generalization of the d=4 formulas to odd dimensions, one can
construct the d=11 invariant of \intthreef\ by 
assuming that
the maximum number of spinor indices of $J_{A_1 ... A_{11}}(x,\t)$ is
$L=9$ and that 
\eqn\highthree{J_{\cu_1 \cu_2 \ua_1 ... \ua_9}(x,\t) = 
(\g_{\cu_1 \cu_2})_{(\ua_1 \ua_2}~~ f_{\ua_3 ... \ua_9 )}(x,\t)}
where 
$f_{\ua_1 ... \ua_9}(x,\t)$ is a superfield satisfying the constraint of
\threef\ that
\eqn\threehigh{\l^\ug\l^{\ua_1} ... \l^{\ua_9} D_\ug f_{\ua_1 ... \ua_9}=0,}
$(\g_{\cu_1\cu_2})_{\ub\ug} =
(\g_{\cu_1\cu_2})_{\ug\ub}$ is the two-form d=11 $\g$-matrix, 
and
$\l^\ua$ is a bosonic spinor satisfying the condition that
$\l\g^\cu\l=0$ for $\cu=0$ to $10$. 

To show that \highthree\ satisfies \covclosed\
in a flat background where the only non-vanishing torsion is 
$T_{\ua\ub}{}^\cu = \g_{\ua\ub}^\cu$, note that \threehigh\ implies that
$ \l^\ug\l^{\ua_1} ... \l^{\ua_9} D_\ug J_{\ua_1 ... \ua_9 \cu_1 \cu_2}=0$.
Since $\l\g^\cu\l=0$, this implies that
\eqn\checkthreehigh{ D_{(\ug} J_{\ua_1 ... \ua_9) \cu_1 \cu_2}=
\g^\bu_{(\ug\ua_1} K_{\ua_2 ... \ua_9) \cu_1 \cu_2 \bu}}
for some 
$K_{\ua_2 ... \ua_9 \cu_1 \cu_2 \bu}$.
If one chooses
$J_{\ua_1 ... \ua_8 \cu_1 \cu_2 \cu_3}$ to be proportional to
$K_{\ua_1 ... \ua_8 \cu_1 \cu_2 \cu_3 }$, one finds
that the first non-trivial constraint of \covclosed\ is satisfied.
Furthermore, the gauge invariance of \gauget\ implies that
$f_{\ua_1 ... \ua_7}$ is defined up to the gauge
transformation 
\eqn\gaugethreefhigh{\d 
f_{\ua_1 ... \ua_7 } =  D_{(\ua_1}
\Sigma_{\ua_2 ... \ua_7)}
+ \g^\cu_{(\ua_1 \ua_2} \Omega_{\ua_3 ... \ua_7) \cu}}
where the $d=11$ $\g$-matrix identity 
$(\g_c)_{(\b_1\b_2} (\g^{c d})_{\b_3\b_4)}=0$ has been used.

As before, components of $J_{A_1 ... A_{11}}$ with more than
two
vector components can be constructed from spinor
derivatives of $J_{\cu_1 \cu_2 \ua_1 ... \ua_9}$
of \highthree\ by using the constraints of \covclosed. In a flat background,
the supersymmetric invariant will therefore have the form
\eqn\supertenj{I = {1\over{11!}}\e^{\cu_1 ... \cu_{11}}\int d^{11} x  ~
J_{\cu_1 ... \cu_{11}}(x,\t=0) =
\int d^d x  \int (d^9 \t)_{\ud_1 ... \ud_9} 
~f_{\ua_1 ... \ua_7}}
where the index contractions need to be worked out.

As in the other cases,
the gauge invariance of \gaugethreefhigh\
implies that \supertenj\ only depends on the $\g$-matrix
traceless components of $f_{\ua_1 .. \ua_7}$
and is invariant under the BRST-trivial deformation
\eqn\gafthree{\d (\l^{\ua_1} ... \l^{\ua_7} f_{\ua_1 ... \ua_7})
 = Q_{d=11}
( \l^{\ua_2} ... \l^{\ua_7} \Sigma_{\ua_2 ... \ua_7}).}
Since the unique state with seven
$\l$'s 
in the BRST cohomology is \topthree, \supertenj\ selects out
the component of 
$\l^{\ua_1} ... \l^{\ua_7} f_{\ua_1 ... \ua_7}$
proportional to \topthree, and is therefore
proportional to 
\eqn\intthreehighf{
T^{((\ua_1 ...\ua_7))[\ud_1 ... \ud_9]}
\int d^{11} x \int (d^9 \t)_{\ud_1 ... \ud_9} ~
f_{\ua_1 ... \ua_7}(x,\t) }
of \intthreef.

To generalize this $d=11$ supersymmetric invariant in a curved
supergravity background, one first defines 
$J_{\cu_1 \cu_2 \ua_1 ... \ua_9}(x,\t) = 
(\g_{\cu_1 \cu_2})_{(\ua_1 \ua_2} f_{\ua_3 ... \ua_9)}(x,\t)$ as in \highthree\
where $D_\ub$ of \threehigh\ is now the spinor derivative in a curved
background. One then computes the other components of
$J_{A_1 ... A_{11}}$ in terms of $f_{\ua_1 ... \ua_7}(x,\t)$ 
using the constraints
of \covclosed. Finally, one plugs the $\t=0$ components of 
$J_{A_1 ... A_{11}}$ into the supersymmetric invariant  
\eqn\superteni{I = {1\over{11!}}\e^{\um_1 ... \um_{11}}\int d^{11} x  ~
e_{\um_{11}}^{A_{11}}(x) ... e_{\um_{1}}^{A_{1}}(x)~J_{A_1 ... A_{11}}(x,\t=0)}
where $e_\um^\au(x)$ and $e_\um^\ua(x)$ are the eleven-dimensional vielbein and
gravitino.

\vskip 20pt
{\bf Acknowledgements:} I would like to thank Rafael Lopes de S\'a
for useful discussions,
CNPq grant 300256/94-9
and FAPESP grant 04/11426-0 for partial financial
support, and the Funda\c c\~ao Instituto de F\'{\i}sica Te\'orica
for their hospitality.

\listrefs
\end